\documentclass[12pt,epsf]{article}
\usepackage{amssymb,amsmath}
\usepackage{graphicx}
\newcommand{\be}{\begin{equation}}
\newcommand{\ee}{\end{equation}}
\newcommand{\bea}{\begin{eqnarray}}
\newcommand{\eea}{\end{eqnarray}}
\newcommand{\beas}{\begin{eqnarray*}}
\newcommand{\eeas}{\end{eqnarray*}}
\newcommand{\ba}{\begin{array}}
\newcommand{\ea}{\end{array}}

\newcommand{\nbox}{{\,\lower0.9pt\vbox{\hrule \hbox{\vrule height 0.2 cm \hskip 0.19 cm \vrule height 0.2 cm}\hrule}\,}}

\def\href#1#2{#2}

\begin{document}
\begin{titlepage}
\hfill
\vbox{
    \halign{#\hfil         \cr
           } % end of \halign
      }  % end of \vbox
\vspace*{20mm}
\begin{center}
{\Large \bf Baryon charge from embedding topology \\ and a continuous meson spectrum \\ in a new holographic gauge theory}

\vspace*{15mm}
\vspace*{1mm}
Mark Van Raamsdonk and Kevin Whyte

\vspace*{1cm}

{Department of Physics and Astronomy,
University of British Columbia\\
6224 Agricultural Road,
Vancouver, B.C., V6T 1W9, Canada}

\vspace*{1cm}
%%\maketitle
\end{center}

\begin{abstract}

We study a new holographic gauge theory based on probe D4-branes in the background dual to D4-branes on a circle with antiperiodic boundary conditions for fermions. Field theory configurations with baryons correspond to smooth embeddings of the probe D4-branes with nontrivial winding around an $S^4$ in the geometry. As a consequence, physics of baryons and nuclei can be studied reliably in this model using the abelian Born-Infeld action. However, surprisingly, we find that the meson spectrum is not discrete. This is related to a curious result that the action governing small fluctuations of the gauge field on the probe brane is the five-dimensional Maxwell action in Minkowski space despite the non-trivial embedding of the probe brane in the curved background geometry.

\end{abstract}

\end{titlepage}

\vskip 1cm

\section{Introduction}

Gauge-theory / gravity duality \cite{maldacena} provides a powerful tool to construct and study strongly coupled field theory systems. In recent years, the set of field theories constructed and analyzed in this way has grown to include examples which are qualitatively similar to systems of great physical interest, including QCD (see e.g. \cite{wittenthermal,kk,ks,beegk,myers,ss1}), superconductors , superfluids, quantum Hall systems, and cold atom systems (see \cite{hartnoll, herzog} for recent reviews of applications to condensed matter systems). While it may be too optimistic to expect that we will be able to find gravitational systems that are exactly dual to real QCD or specific real-world condensed matter systems, these model systems can provide significant qualitative insight into generic phenomena that arise in strongly coupled systems similar to the real-world examples. There are already examples (e.g. the very low viscosity to entropy ratio for the quark-gluon plasma produced in heavy ion collisions) where the insight gained from holographic models offers the best theoretical understanding of an experimentally measured phenomenon (see e.g. \cite{kss,sonstarinets,gk}).

With such potential for new theoretical insight into physically interesting systems, it seems fruitful to explore a wide variety of holographically constructed field theories. In doing so, we may uncover new qualitative phenomena in strongly coupled field theories that could help explain real-world physical phenomena or, more generally, lead to an improved understanding of quantum field theory at strong coupling. In addition, amassing a large number of detailed examples will help reveal which features of these systems are generic (and thus more likely to apply to other systems for which we may not have a precise gravity dual), and which are peculiar to specific constructions.

Motivated by these considerations, we study in this paper a new holographic field theory closely related to the Sakai-Sugimoto model \cite{ss1} of holographic QCD. Specifically, our theory has the same adjoint sector, but a different fundamental sector since we use probe D4-branes instead of probe D8-branes. Thus, our model is based on a brane construction where both the ``color branes'' (which give rise to the adjoint sector) and the ``flavor branes'' are D4-branes. The relative orientation for the two sets of branes is:
\beas
\ba{rcccccccccc}
&0&1&2&3&4&5&6&7&8&9 \cr
N_c \; D4 &\times &\times&\times&\times&\times&&&&&\cr
N_f \; D4 &\times &\times&\times&\times&&\times&&&&
\ea
\eeas
as shown in figure 1.
\begin{figure}
\centering
\includegraphics[width=0.3\textwidth]{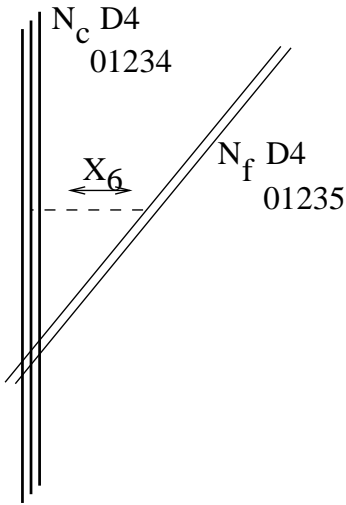}
\caption{Brane construction for the holographic field theory. }
\end{figure}
At weak coupling, this system has a (complex scalar) tachyon in its low-energy spectrum coming from the D4-D4' strings. However, by separating the branes in a transverse direction (e.g. the 6 direction), we can arrange for this tachyon to become massless or massive. The lightest D4-D4' fermions have string scale masses in this case. For the original theory with no transverse separation, the geometrical $SO(5)$ symmetry present in the adjoint sector of the theory is broken to $SO(4)$, while the theory defined with a transverse separation between the two sets of branes has this symmetry broken to $SO(3)$.

In order to obtain a decoupled field theory, we want to take a low-energy decoupling limit in this brane setup. We would like to do this in such a way that the lightest modes of the D4-D4' strings survive. It is plausible that we can tune the transverse separation of the two sets of branes as we take the limit to achieve this. In practice, we do not actually define an explicit decoupling limit starting from a brane configuration in asymptotically flat space. Instead, we do something much simpler (following the Sakai-Sugimoto example). We will always consider the limit where $N_f \ll N_c$, so that the fundamental matter does not affect the physics of the adjoint sector (i.e. the quenched approximation is accurate). Then the addition of fundamental matter can be achieved simply by adding probe D4-branes into the geometry dual to the field theory describing the low-energy degrees of freedom of the $N_c$ D4-branes. Thus, instead of starting with color and flavor branes in asymptotically flat space, we just look for a stable configuration of probe D4-branes in the geometry dual to the color branes such that the configuration preserves the desired symmetries.

Since we are working in a consistent background of string theory, we can say that the model we describe is some fully consistent quantum field theory sharing many qualitative features with QCD. Based on the weak coupling picture, it is tempting to suggest that the model we describe is one where the fundamental quarks are purely scalar, since the D4-D4' fundamental fermions have string scale masses when the branes are tuned to make the lightest scalar massless or massive. Achieving a model with scalar quarks was one of the original motivations for studying this model, since we were interested to look at the qualitative similarities and differences between our model and the Sakai-Sugimoto model (where the fundamental matter is fermionic), and also to see whether any new qualitative phenomena appear in the physics of strongly coupled fundamental bosons. However, since our actual construction is less direct than an explicit decoupling limit, we cannot say definitively that the model includes only scalar quarks.

\vskip 0.2 in
\noindent
{\bf Outline and Summary}
\vskip 0.1 in
\noindent
After a review of the basic setup in section 2, we carry out the analysis of probe brane embeddings in section 3, focusing on the case $N_f=1$. We find a two-parameter family of D4-brane embeddings.\footnote{A holographic field theory corresponding to a different class of probe D4-brane embeddings was studied in \cite{gp} among other examples. Another model with D4-brane probes in a different background was studied in \cite{cps}.} These are labeled by a parameter $y_0$ that measures how far into the IR of the geometry the probe brane reaches and a parameter $v_0$ that controls how much the brane is tilted into the compact direction of the field theory. The embeddings in this family correspond to the vacuum solutions for a two-parameter family of field theories. In each case, the solution preserves $SO(3) \sim SU(2)$ global symmetry, but for a one parameter family, this symmetry arises via a spontaneous breaking from $SO(4) \sim SU(2) \times SU(2)$. For this special case, there is a family of solutions with the same $SO(4)$-preserving asymptotics. Each of the solutions preserves only $SO(3)$, so we would expect an $SO(3)$ vector of massless scalar goldstone bosons associated with the broken symmetry. However, we do not find an ordinary discrete spectrum of mesons as in other models, but rather a continuous spectrum (as we would have in a conformal field theory)\footnote{A continuous meson spectrum was found also in \cite{Arean:2006vg} for a defect theory, but it is not clear whether the underlying mechanism is the same.}. This is related to the fact that the action governing small fluctuations of the gauge field on the probe brane is the ordinary Maxwell action in {\it Minkowski space} despite the nontrivial embedding of the probe brane in the curved background geometry. We do not have a good interpretation for this from the field theory point of view. However, the result crucially depends on the relative normalization of the Chern-Simons and Born-Infeld terms in the brane action; if this normalization is changed at all, we get an ordinary discrete spectrum of mesons. Note that our results refer only to the small fluctuation analysis (keeping quadratic terms in the action for fluctuations of the probe brane about its equilibrium configurations). It is possible that including the effects of higher-order terms in the Born-Infeld action may have interesting effects (in particular, they break the ``accidental'' five-dimensional Lorentz invariance present in the quadratic action for the gauge field), but this analysis is beyond the scope of the present work.

One of the most interesting feature of our model that the baryonic sector can be studied very reliably, as we discuss in section 5. As in the Sakai-Sugimoto model, baryon charge arises from D4-branes which wrap an $S^4$ in the geometry (see section 5 for a review). But in our D4-D4 system, these wrapped D4-branes can smoothly reconnect with the probe D4-branes of the original embedding. Thus, states in the field theory with nonzero baryon number correspond to smooth D4-brane embeddings of different topology from the vacuum embedding. Baryon charge in the field theory corresponds in the bulk to a topological charge $\pi_4(S^4)$ for the embedding (relative to the original embedding). These smooth embeddings can be studied reliably using the abelian Born-Infeld action, so properties such as baryon mass and nuclear binding energies should be under complete control in the model. This is in contrast to the Sakai-Sugimoto model, where baryons are described by instanton-like configurations of the Yang-Mills fields on the probe D8-branes whose size is string-scale. In that case, higher $\alpha'$ corrections to the Born-Infeld action should be important for a completely reliable treatment of baryon physics.\footnote{There have nevertheless been many studies of baryons and baryon physics in the Sakai-Sugimoto model making use of various approximations, see e.g. \cite{baryons} for some of the early work.} On the other hand, given the results for the meson sector, our model is clearly much further from real QCD than the Sakai-Sugimoto model.

\section{Setup}

In this section, we review the basic construction of our holographic field theory. We begin by describing the adjoint sector, and then describe the addition of flavor fields via the embedding of probe branes in the dual geometry.

\subsection{Adjoint sector}

The adjoint sector of our model was originally proposed by Witten \cite{wittenthermal} as a construction of non-supersymmetric Yang-Mills theory. It is defined by the low-energy decoupling limit of $N \equiv N_c$ D4-branes wrapped on a circle of length $2 \pi R$ with anti-periodic boundary conditions for the fermions. This part of the theory has two dimensionless parameters, $N_c$ and a coupling constant
\[
\lambda = {\lambda_{D4} \over 2 \pi R} \; ,
\]
where
\[
\lambda_{D4} = g^2_{YM} N \; .
\]
The dimensionless parameter $\lambda$ is the effective four-dimensional coupling at the Kaluza-Klein scale. For small $\lambda$, this coupling runs to strong coupling at a smaller scale
\[
\Lambda_{QCD} \sim {1 \over R} e^{- c \over \lambda}
\]
where the physics should be exactly that of pure 3+1 dimensional Yang-Mills theory (thanks to fermion masses generated by the antiperiodic boundary conditions and scalar masses generated at one loop). For large $\lambda$, the dual gravity theory becomes weakly curved, and physics is well described by type IIA supergravity on a background
\bea
ds^2 &=& \left({U \over R_4} \right)^{3 \over 2}(\eta_{\mu \nu} dx^{\mu} dx^{\nu} + f(U) dx_4^2) + \left({R_4 \over U} \right)^{3 \over 2}({1 \over f(U)} dU^2 + U^2 d \Omega_4^2)  \cr
e^{\phi}  &=& g_s \left({U \over R_4} \right)^{3 \over 4} \cr
F_4 &=& {(2 \pi)^3 N_c (\alpha')^{3 \over 2} \over \omega_4} \epsilon_4 \; .
\label{metric}
\eea
Here $\omega_4 = {8 \over 3}\pi^2$ is the volume of a unit 4-sphere, $\epsilon_4$ is the volume form on $S^4$, and
\be
\label{ef}
f(U) = 1 - \left({U_0 \over U} \right)^3 \; .
\ee
The $x_4$ direction, corresponding to the Kaluza-Klein direction in the field theory, is taken to be periodic, with coordinate periodicity $2 \pi R$, however, it is important to note that this $x_4$ circle is contractible in the bulk since the $x_4$ and $U$ directions form a cigar-type geometry.

The parameters $R_4$ and $U_0$ appearing in the supergravity solution are related to the string theory parameters by
\[
R^3_4 = \pi g_s N l_s^3 \qquad \qquad U_0 = {4 \pi \over 9 R^2} g_s N l_s^3
\]
while the four-dimensional gauge coupling $\lambda$ is related to the string theory parameters as
\[
\lambda = 2 \pi {g_s N l_s \over R} \; .
\]
In terms of the field theory parameters, the dilaton and string-frame curvature at the tip of the cigar (the IR part of the geometry) are of order $\lambda^{3 \over 2}/N$ and $\sqrt{\lambda}$, so as usual, supergravity will be a reliable tool for studying the infrared physics when both $\lambda$ and $N$ are large (in this case, with $N >> \lambda^{3 \over 2}$).

\subsection{Fundamental matter}

The addition of fundamental matter manifests itself through the appearance of $N_f$ probe D4-branes in the geometry dual to the adjoint sector. These D4-branes are extended along the $x^\mu$ directions, and are described by a one-dimensional path in the remaining radial, sphere, and $x_4$ directions. It is convenient to redefine coordinates so that the metric in the radial and sphere directions takes the form
\be
\label{rhoth}
\alpha(\rho)(d\rho^2 + \rho^2 d \Omega_4^2)
\ee
These coordinates should satisfy
\[
{dU \over U \sqrt{f(U)}} = {d \rho \over \rho}
\]
From this, we find the map
\[
{U \over U_0} = \left( {1 \over 2} \left( {\rho \over \rho_0} \right)^{3 \over 2} + {1 \over 2} \left( {\rho_0 \over \rho} \right)^{3 \over 2} \right)^{2 \over 3} \; .
\]
where $\rho_0 = U_0 2^{-{2 \over 3}}$. Locally, the metric (\ref{rhoth}) is conformally equivalent to $R^5$, however we should note that the space has an infrared end at $\rho = \rho_0$ where the $X^4$ circle contracts to a point. Thus, the ball $\rho < \rho_0$ is not part of the geometry.

The equilibrium brane configurations come in from radial infinity, reach some minimum value of the radial coordinate, and go back out to radial infinity. These configurations asymptote to two specific directions on the $S^4$. The boundary conditions generically break the $SO(5)$ symmetry to $SO(3)$, and we expect that the minimum action embeddings do not break the symmetry further. In other words, we expect that the stable configurations will lie in a single plane in the $R^5$ appearing in  (\ref{rhoth}), so it will sometimes be convenient to use coordinates
\[
d\rho^2 + \rho^2 d \Omega_4^2 = dr^2 + r^2 d \theta^2 + d\vec{x}_T^2
\]
in terms of which the equilibrium D4-brane configurations will be specified by $\vec{x}_T = 0 $ and $r(\theta) = \rho(\theta)$ (note that $\rho=r$ for $\vec{x}_T = 0 $).

To write the action for the probe D4-branes, we focus on the case of a single brane, for which we can use the abelian Born-Infeld action
\be
\label{dbi}
S = - \mu_4 \int d^{5} \sigma e^{-\phi} \sqrt{-\det(g_{ab} + \tilde{F}_{ab})}
\ee
together with the Chern-Simons part:
\be
\label{CS}
S = \mu_4 \int \sum C \wedge e^{\tilde{F}}
\ee
where
\[
\tilde{F} = 2 \pi \alpha' F \; .
\]
We choose static gauge $X^\mu = \sigma^\mu$ for the field theory directions, and describe the nontrivial part of the embedding by functions $X^4(\sigma),r(\sigma), \theta(\sigma), X^T_i(\sigma)$, where $\sigma$ parameterizes the remaining coordinate along the brane. The pull-back metric appearing in the Born-Infeld action is then given explicitly by
\beas
g_{\mu \nu} &=& G_{\mu \nu} + G_{44} \partial_\mu X_4 \partial_\nu X_4 +G_{rr} \partial_\mu r \partial_\nu r + G_{\theta \theta} \partial_\mu \theta \partial_\nu \theta + G_{ij} \partial_\mu X^i_T \partial_\nu X^j_T \cr
g_{\mu \sigma} &=& G_{44} \partial_\mu X_4 \partial_\sigma X_4 +G_{rr} \partial_\mu r \partial_\sigma r + G_{\theta \theta} \partial_\mu \theta \partial_\sigma \theta + G_{ij} \partial_\mu X^i_T \partial_\sigma X^j_T \cr
g_{\sigma \sigma} &=& G_{44} \partial_\sigma X_4 \partial_\sigma X_4 +G_{rr} \partial_\sigma r \partial_\sigma r + G_{\theta \theta} \partial_\sigma \theta \partial_\sigma \theta + G_{ij} \partial_\sigma X^i_T \partial_\sigma X^j_T \cr
\eeas
For now, we are interested in equilibrium brane configurations, which we assume have $X_T^i=0$, so we keep only terms in the action involving $X_4(\sigma)$, $r(\sigma)$ and $\theta(\sigma)$. With this simplification, the Born-Infeld part of the action becomes
\be
S = - {\mu_4 \over g_s}  \int d \sigma d^4 x \; H(r(\sigma)) \sqrt{r^2 \left( {d\theta \over d \sigma}\right)^2  + \left({d r \over d \sigma }\right)^2 }
\ee
where
\[
H(r) = {r^{3 \over 2} \over R^{3 \over 2}} \left(1 + { \rho_0^3 \over r^3}\right)^{5 \over 3}
\]

We now turn to the Chern-Simons part of the action. Since the background we are considering involves a non-zero Ramond-Ramond four-form flux, the potentials $C_3$ and the dual $C_5$ are non-zero. For the configurations that we are considering (which are translation-invariant in the field theory directions), the pull-back of $C_3$ is zero, but we have a non-zero pull-back for $C_5$. We find (see appendix A for a derivation):
\[
C_5 = {\pi N (\alpha')^{3 \over 2} \over R_4^6} (U^3 - U_0^3) dt \wedge dx_1 \wedge dx_2 \wedge dx_3 \wedge dx_4 \; ,
\]
so the Chern-Simons term in the action is
\beas
S_{CS} &=& \mu_4 \int C_5 \cr
&=& \mu_4 {\pi N (\alpha')^{3 \over 2} \over R_4^6} \int d \sigma d^4 x \; (U^3 - U_0^3) {\partial X_4 \over \partial \sigma}
\eeas

\section{Vacuum solutions}

For our calculations of the vacuum configurations, it is convenient to fix the remaining reparametrization invariance by choosing $\sigma = \theta$. If we also define
\[
y = {r \over \rho_0} \qquad \qquad x = \sqrt{ \rho_0 \over R_4^3} X_4\; ,
\]
the resulting action is
\beas
S &=& {\mu_4 \over g_s}  {\rho_0^{5 \over 2} \over R_4^{3 \over 2}} \left\{ -\int d \theta h(y) \sqrt{y^2  + (y')^2 + g(y)(x')^2} + \int d \theta q(y) x' \right\}
\eeas
where
\beas
h(y) &=& y^{3 \over 2} (1 + {1 \over y^3})^{5 \over 3} \cr
g(y) &=& y {(y^3 - 1)^2 \over (y^3 + 1)^{4 \over 3}} \cr
q(y) &=& {(y^3 - 1)^2 \over y^3}
\eeas
Since the resulting Lagrangian density does not depend explicitly on $\theta$, we have a $\theta$-independent quantity (analogous to energy for a time-independent Lagrangian density) given by
\[
y' {\partial S \over \partial y'} + x' {\partial S \over \partial x'} - S = {hy^2 \over \sqrt{y^2  + (y')^2 + g(y)(x')^2}}
\]
Since the geometry caps off smoothly at some finite value of $y$, smooth brane configurations must have some minimal value of $y$ for which $y'=0$. Calling this value $y_0$, and calling the derivative $x'$ at this point $v_0$, we have
\[
{h(y) y^2 \over \sqrt{y^2  + (y')^2 + g(y)(x')^2}} = {h(y_0) y_0^2 \over \sqrt{y_0^2 + g(y_0)v_0^2}} \equiv B
\]
The action also does not depend explicitly on $x$ (only on $x'$), so we have another constant
\[
{-h(y) g(y) x' \over \sqrt{y^2  + (y')^2 + g(y)(x')^2}} + q(y) = q(y_0) - {h(y_0) g(y_0) v_0 \over \sqrt{y_0^2 + g(y_0)v_0^2}} \equiv C \; .
\]
From the equations above, we can eliminate $x$ to get:
\be
\label{dydth}
{dy \over d \theta} = \pm y \sqrt{{y^2 \over B^2}(h^2(y) - {(q(y)-C)^2 \over g(y)}) - 1} \; .
\ee
We also find:
\be
\label{dxdy}
{ dx \over dy} = \pm {y (q(y) - C) \over B g(y) \sqrt{{y^2 \over B^2}(h^2(y) - {(q(y)-C)^2 \over g(y)}) - 1}}
\ee
These two equations can be integrated to find $\theta(y)$ and $x(y)$.

Integrating, we find
\be
\label{soln}
\theta(y) =  \int_{y_0}^y  d\tilde{y} {1 \over \tilde{y} \sqrt{{\tilde{y}^2 \over B^2}(h^2(\tilde{y}) - {(q(\tilde{y})-C)^2 \over g(\tilde{y})}) - 1}}\; .
\ee
where we define $\theta=0$ to be the angle at which the brane embedding reaches its minimum value of $y$. From this expression, it is straightforward to check that for any value $y_0 > 1$ and $v_0$,\footnote{Recall that y=1 represents the IR end of the geometry.} $\theta$ approaches a finite value as $y$ goes to infinity.

\begin{figure}
\centering
\includegraphics[width=0.75\textwidth]{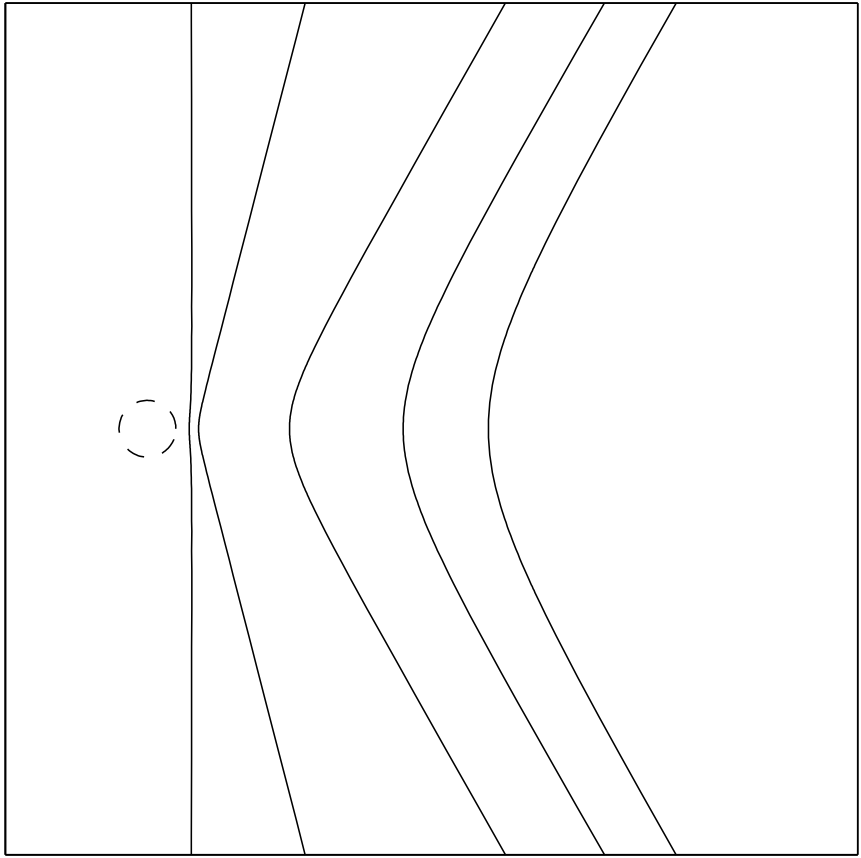}
\caption{Examples of brane embeddings (in the $y-\theta$ plane) for $v_0=0$. The asymptotic angle between the two ends of the brane is $2 \theta_\infty$, which ranges from $\pi$ for the stable embedding which extends to the smallest values of $y$, down to some value $2 \theta_{Max} \approx 2.088$ in the limit $y_0 \to \infty$.}
\end{figure}

\begin{figure}
\centering
\includegraphics[width=0.75\textwidth]{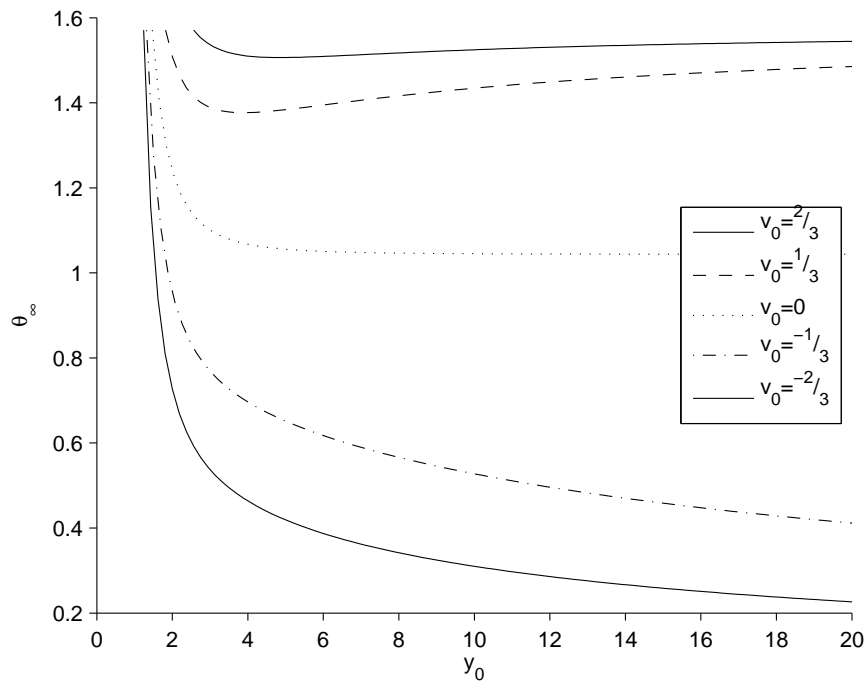}
\caption{Asymptotic angle $\theta_\infty$ on the sphere vs minimum brane position $y_0$ in radial direction, for various values of $v_0$. Angle $\theta$ is defined to be zero at $y=y_0$. }
\end{figure}

Thus, the brane configurations asymptote to lines of constant $\theta$, as shown in figure 2. The relation between $y_0$ and the maximal value of $\theta$ is given by
\be
\label{thmax}
\theta_\infty(y_0) =  \int_{y_0}^\infty  d\tilde{y} {1 \over \tilde{y} \sqrt{{\tilde{y}^2 \over B^2}(h^2(\tilde{y}) - {(q(\tilde{y})-C)^2 \over g(\tilde{y})}) - 1}}\; .
\ee
and plotted in figure 3 for various values of $v_0$. For any given $v_0$, we find that for large $y_0$, the asymptotic angle approaches a limiting value $\theta_{Max}(v_0)$, given by
\[
\theta_{Max}(v_0) = \left\{ \ba{ll} 0 & v_0 <0 \cr 1.044144565 & v_0 = 0 \cr {\pi \over 2} & v_0 > 0 \ea \right.
\]
For each $v_0$, there is a special value $y_0=y_*(v_0)$ for which the two asymptotic ends of the brane go towards diametrically opposite points on the sphere. As $y_0$ approaches 1, $\theta_\infty$ increases without bound, corresponding to brane embeddings that wrap multiple times around the $\theta$ direction. However, for $y_0 < y_*(v_0)$ these embeddings are perturbatively unstable to ``unwrapping,'' i.e. slipping over the spherical hole in the geometry, as seen in figure 4. The perturbative instability will be demonstrated explicitly in the next section.

\begin{figure}
\centering
\includegraphics[width=0.3\textwidth]{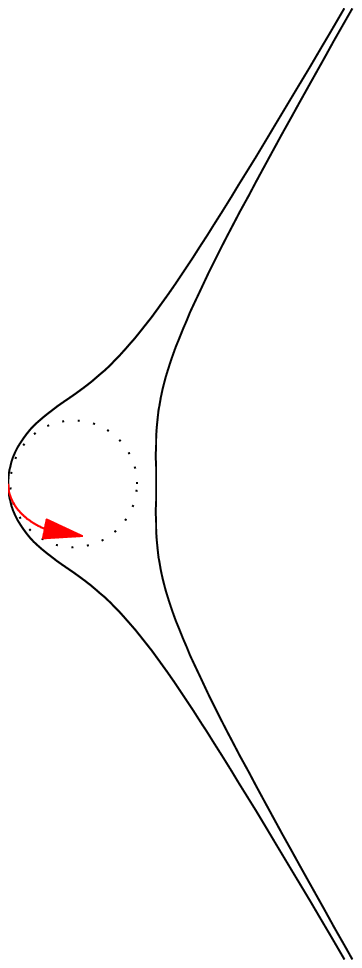}
\caption{Example of multiple embeddings for the same asymptotic sphere angles. Only embeddings which do not ``wrap'' the sphere are stable. The rest are perturbatively unstable to slipping around the sphere, as shown.}
\end{figure}

\begin{figure}
\centering
\includegraphics[width=0.6\textwidth]{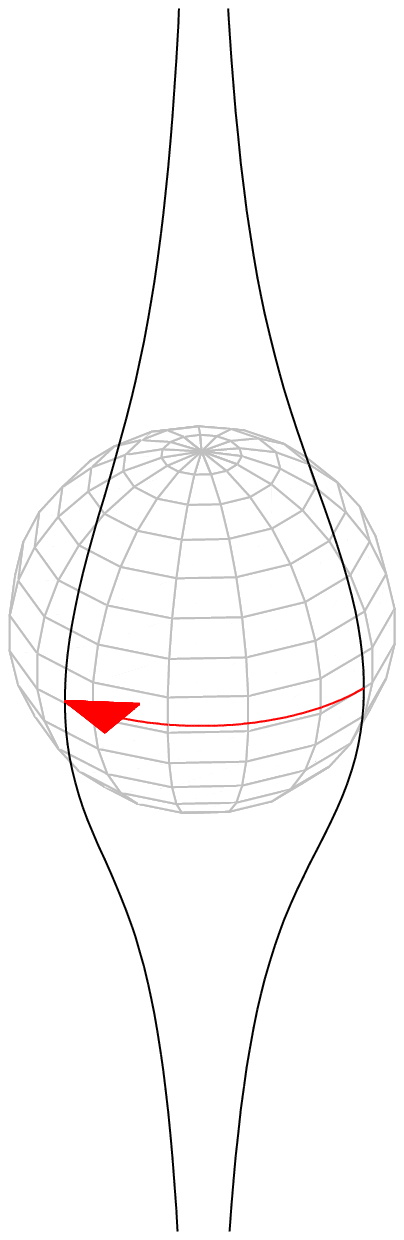}
\caption{Geometrical interpretation of goldstone bosons for the special case $\theta_\infty = \pi/2$.}
\end{figure}

For the special case $\theta_\infty = \pi/2$, the two ends of the probe brane go to diametrically opposite points on the $S^4$. For these asymptotics, we actually have a family of embeddings related by the $SO(4)$ rotations that fix these diametrically opposite points on the sphere as shown in figure 5. This case corresponds to a spontaneous breaking of $SO(4) \to SO(3)$ (equivalently $SU(2) \times SU(2) \to SU(2)$), and we must therefore have an $SO(3)$ vector of massless goldstone bosons associated with the broken symmetry. It is these bosons that become tachyonic if we increase $\theta_\infty$ beyond $\pi/2$ for fixed $v_0$. This is very similar to the naive brane picture in figure 1, where a tachyon develops if the transverse separation between the branes becomes too small.

\begin{figure}
\centering
\includegraphics[width=0.75\textwidth]{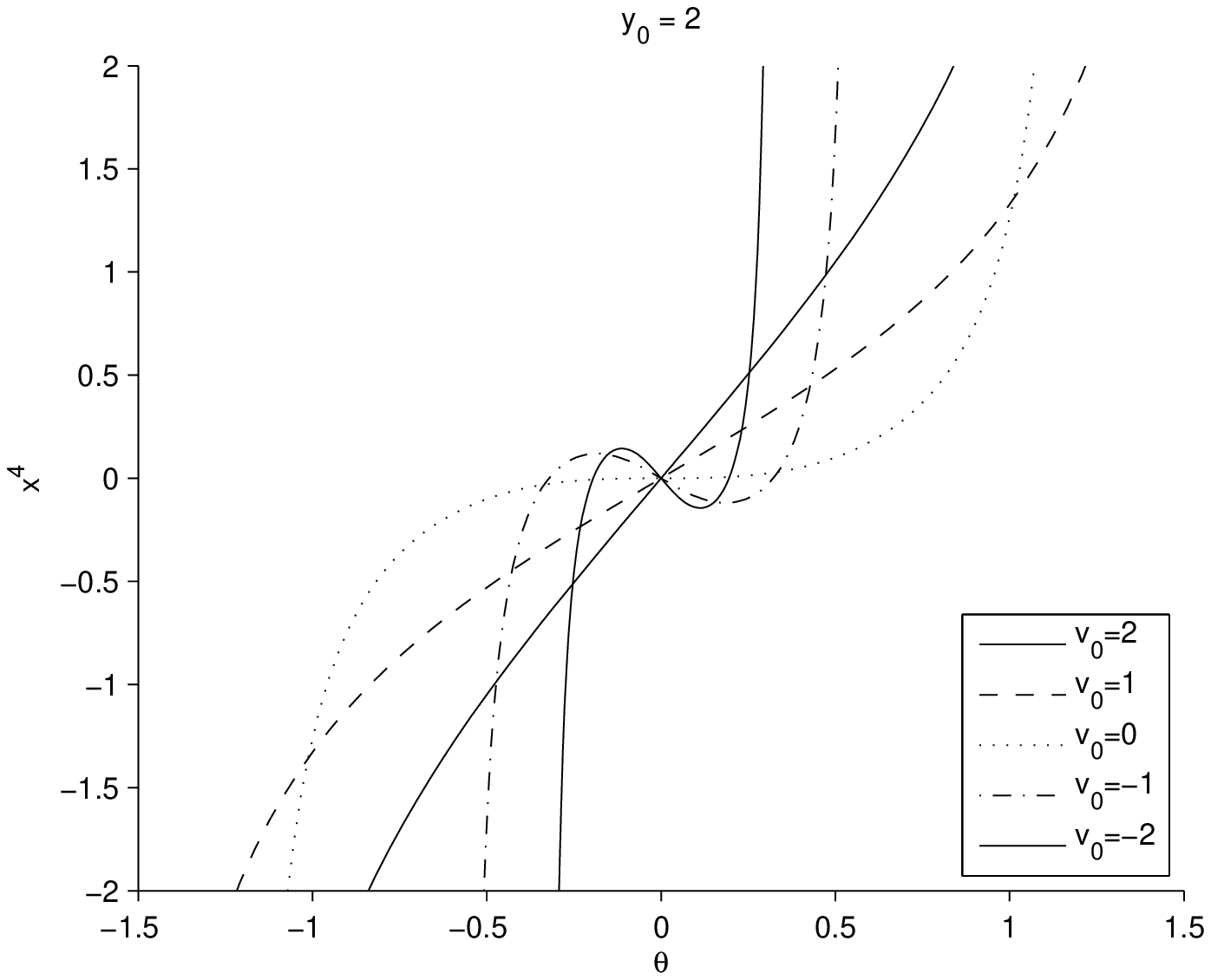}
\caption{Behavior of $x_4$ vs $\theta$ for various values of $v_0$ at $y_0=2$. As a function of $y$, the slope $dx_4/dy$ approaches a constant in each case.}
\end{figure}

The behavior of the embedding in the $X_4$ direction can be obtained by integrating (\ref{dxdy}). We find that for any value of $y_0$, the $x_4(y)$ asymptotes to a constant positive slope $dx/dy$, so that the brane continues to wrap the $x_4$ direction as we go out to $y = \infty$. Because of the Chern-Simons coupling, the probe brane ``prefers'' a positive slope $dx/dy$; we see that the asymptotic slope is positive even if slope $dx_4/d \theta$ is negative at $y=y_0$. The behavior of $x_4$ is shown in figure 6.

\section{Meson spectrum and stability}

In this section, we consider small fluctuations about the equilibrium brane configurations found in the previous section. We would like to determine which of the embeddings are perturbatively stable, and for these embeddings, to determine the spectrum of small fluctuations that gives the meson spectrum for the theory.

To determine the fluctuation spectrum, we start with the brane action (\ref{dbi}) and expand to quadratic order about a chosen solution, parameterized by $(y_0,v_0)$. We consider all possible bosonic fluctuations, which include fluctuations in the $x_4$ direction, fluctuations in the three transverse directions along the sphere (which we label by an $SO(3)$ triplet of scalar fields $X_T$), fluctuations in the $r-\theta$ plane, and the gauge field fluctuations.

In general, for the scalar field modes, the action for small fluctuations about the vacuum solution takes the form
\be
\label{scac}
S = -C \int d^4 x \int_{y_0}^\infty dy \left\{ {1 \over 2} A(y) \left( {\partial \phi \over \partial x^\mu} \right)^2 + {1 \over 2} {\rho_0 \over R^3} B(y) \left( {\partial \phi \over \partial y} \right)^2 + {1 \over 2} {\rho_0 \over R^3} C(y) \phi^2  \right\}
\ee
while for the gauge fields, we have
\[
S = -C \int d^4 x \int_{y_0}^\infty dy \left\{ {1 \over 4} A(y) F_{\mu \nu} F^{\mu \nu} + {1 \over 2} {\rho_0 \over R^3} B(y) F_{\mu y} F^{\mu y}  \right\}
\]
It is convenient to define the functions $A,B$, and $C$ using the function
\[
R(y) = (1 + y^2 {d \theta_0 \over dy}^2 + g(y){dx_0 \over dy}^2)^{-{1 \over 2}} \; .
\]
where ${d \theta_0 \over dy}$ and ${dx_0 \over dy}$ refer to the background embedding functions and are given in terms of $y$, $y_0$, and $v_0$ by the equations (\ref{dydth}) and (\ref{dxdy}). Explicitly, we have
\[
R(y) = \sqrt{1 - {B^2 \over y^2 h^2(y)} - {(q(y)-C)^2 \over h^2(y) g(y)}} \; .
\]
where $B(y_0,v_0)$ and $C(y_0,v_0)$ are defined in the previous section. In the special case where $v_0 = 0$, we get:
\[
R(y) = \sqrt{1-{y^5(y_0^3+1)^{10 \over 3} \over y_0^5(y^3+1)^{10 \over 3}} - {(y^3-1)^2 \over (y^3+1)^2}\left(1-{ y^3 (y_0^3-1)^2 \over y_0^3 (y^3-1)^2  } \right)^2} \; .
\]
For the transverse scalar ($X_T$) fluctuations, we find
\beas
A(y) &=&  {(y^3 + 1) \over  y^{9 \over 2}} R^{-1}(y) \cr
B(y) &=&  {(y^3 + 1)^{5 \over 3} \over  y^{7 \over 2}} R(y) \cr
C(y) &=&  {1 \over 2} {(y^3 + 1)^{2 \over 3} \over  y^{11 \over 2}} R(y) \left\{ (3 y^3 -7)(1+y^2{d \theta_0 \over dy}^2) + {6 y (y^3-1)(y^6+1) \over (y^3 + 1)^{4 \over 3} } {dx_0 \over dy}^2 \right\} \cr
&& \qquad - 3 {dx_0 \over dy} {(y^3 + 1)(y^3 -1) \over y^5}
\eeas
where the last term in $C$ comes from the Chern-Simons action.

For the gauge field fluctuations, we find
\beas
A(y) &=& {1 \over y^{1 \over 2} (1+y^3)^{1 \over 3}} R^{-1}(y) \cr
B(y) &=&  y^{1 \over 2} (1+y^3)^{1 \over 3} R(y)
\eeas
Finally, the $\theta$ and $X_4$ fluctuations mix with each other, and the fluctuation action for the combination $(\theta, x)$ is given as above where now $A$ and $B$ are matrices
\beas
A(y) &=& R(y) \left( \ba{cc} {(y^3 + 1) \over y^{5 \over 2}} (1 + g(y){dx_0 \over dy}^2) & - {dx_0 \over dy} {d \theta_0 \over dy} {(y^3-1)^2 \over y^{3 \over 2} (y^3 + 1)^{1 \over 3}} \cr
- {dx_0 \over dy} {d \theta_0 \over dy} {(y^3-1)^2 \over y^{3 \over 2} (y^3 + 1)^{1 \over 3}} &  {(y^3-1)^2 \over y^{7 \over 2} (y^3 + 1)^{1 \over 3}}(1 + y^2 {d \theta_0 \over dy}^2) \ea \right) \cr
B(y) &=& R^3(y) \left( \ba{cc} {(y^3 + 1)^{5 \over 3} \over y^{3 \over 2}} (1 + g(y){dx_0 \over dy}^2) & - {dx_0 \over dy} {d \theta_0 \over dy} {(y^3-1)^2   (y^3 + 1)^{1 \over 3} \over y^{1 \over 2}} \cr
- {dx_0 \over dy} {d \theta_0 \over dy} {(y^3-1)^2   (y^3 + 1)^{1 \over 3} \over y^{1 \over 2}} &  {(y^3 + 1)^{1 \over 3} (y^3-1)^2 \over y^{5 \over 2} }(1 + y^2 {d \theta_0 \over dy}^2) \ea \right) \cr \; .
\eeas

In this special cases $v_0 = \pm \infty$, we have ${d \theta_0 \over dy} = 0$, and so these matrices become diagonal.

\subsection{Scalar modes}

For the scalar modes, the fluctuation actions above give rise to an equation of motion
\[
-A(y) {\partial^2 \phi \over \partial x^2_\mu}  - {\rho_0 \over R^3} {\partial  \over \partial y} \left( B(y) {\partial \phi \over \partial y}\right) +  {\rho_0 \over R^3} C(y) \phi = 0
\]

We look for solutions of the form
\[
\phi(x,y) = e^{i k \cdot x} f(y)
\]
where $f(y)$ falls off fast enough so that the integral over $y$ in the action converges (i.e. so that $\phi$ is a normalizible fluctuation). With this ansatz, the equation reduces to
\be
\label{Schrod}
- {\rho_0 \over R^3} {\partial \over \partial y} \left( B(y) {\partial f \over \partial y}\right) +  ( {\rho_0 \over R^3} C(y) - \lambda A(y) ) f = 0
\ee
where $\lambda = m^2$ represents the four-dimensional mass of the fluctuation.

\subsection{Gauge modes}

In order to solve the gauge field fluctuation equations, it is convenient to choose a gauge
\[
\partial_\mu A^\mu + A^{-1} \partial_y (B A_y) = 0 \; .
\]
With this choice, the equations of motion for the various components of $A$ decouple. For the components in the field theory directions, we have
\be
\label{amu}
\partial^2 A^\nu + A^{-1} \partial_y (B \partial_y A^\nu) = 0
\ee
while for the $y$ component, we have
\be
\label{ay}
\partial^2 A_y + \partial_y( A^{-1} \partial_y (B A_y)) = 0 \; .
\ee
This set of equations has a residual gauge invariance under transformations
\be
\label{GI}
A_\mu \to A_\mu + \partial_\mu \lambda \qquad A_y \to A_y + \partial_y \lambda
\ee
for any $\lambda$ satsfying
\be
\label{lambda}
\partial^2 \lambda + A^{-1} \partial_y (B \partial_y \lambda) \; .
\ee
This allows us to make a further gauge choice $A_y=0$. The gauge field fluctuation modes are then captured by solutions to the equation (\ref{amu}). These can be found by separation of variables, considering solutions of the form
\[
A_\mu  = \epsilon_\mu(k) e^{i k \cdot x} a(y)
\]
where we require
\[
k \cdot \epsilon(k) = 0
\]
by our original gauge condition, and where $a(y)$ is a normalizible solution to
\[
-k^2 A(y) a +  \partial_y (B \partial_y a) \; .
\]
This eigenvalue equation is the same type (\ref{Schrod}) as we obtain from the scalar equation.

\subsection{Converting to a quantum mechanics problem}

For both gauge and scalar modes, we need to determine the values of $\lambda$ for which normalizible solutions to the equation (\ref{Schrod}) exist. We can convert this into a simple quantum mechanics problem as follows. First, note that the equation arises from an action
\be
S = \int_{y_0}^\infty dy \left\{ {1 \over 2} B(y) \left( {\partial \phi \over \partial y} \right)^2 + {1 \over 2}  (C(y) - \tilde{\lambda} A(y)) \phi^2  \right\}
\ee
where we have defined $\tilde{\lambda} = R^3 \lambda / \rho_0$. Now, we define a new variable $z$ such that
\[
z = \int_{y_0}^y {d\tilde{y} \over B(\tilde{y})}
\]
such that
\[
{dy \over dz} = B(y) \qquad \qquad z(y_0) = 0\; .
\]
In the new variables, the action becomes
\be
S = \int_0^{z_\infty} dz \left\{ {1 \over 2} \left( {\partial \phi \over \partial z} \right)^2 + {1 \over 2}  B(y(z))(C(y(z)) - \tilde{\lambda} A(y(z))) \phi^2  \right\}
\ee
where
\[
z_\infty = \int_{y_0}^\infty {d\tilde{y} \over B(\tilde{y})}
\]

This gives rise to the time-independent Schrodinger equation for $E=0$,
\[
-f''(z) + V(z) f(z) = 0 \; ,
\]
so our problem is reduced to determining for which values of $\tilde{\lambda}$ the Schrodinger equation with potential
\be
\label{veff}
V(z) =  B(y(z))(C(y(z)) - \tilde{\lambda} A(y(z)))
\ee
has a bound state with zero energy.

\subsection{Gauge field fluctuations: a continuous spectrum}

We consider first the gauge field fluctuations. Here, we note that $C=0$ and $A(y)B(y) = 1$, so our quantum mechanics potential is simply
\[
V_A(z) = - \lambda \; .
\]
Also, in this case, we find $z_\infty = \infty$. So we do not have any bound states for any $\lambda$, though there are zero-energy solutions
\[
f_\lambda(z) = e^{\pm i \sqrt{\lambda} z}
\]
to the Schrodinger equation for any $\lambda \ge 0$. These do not fall off fast enough at large $z$ to be normalizible, but we can superpose solutions with different $\lambda$ to get normalizible solutions. Since the four-momentum in the field theory directions is related to $\lambda$ by $-k^2 = \lambda$, these superpositions will not be eigenstates of four-momentum. Thus there are no true particle states in the field theory arising from the gauge mode fluctuations.

To understand this better, we note that with the new radial variable, the action governing small fluctuations of the gauge field is exactly the Maxwell action in $4+1$ dimensional Minkowski space (despite the nontrivial embedding of the brane in a non-trivial curved space!)
\[
S \propto \int d^4 x d z \{ - {1 \over 4} F_{AB} F^{AB} \}
\]
Configurations with finite energy in the field theory correspond to solutions of these $5D$ Maxwell equations that fall off sufficiently rapidly for large $z$ and $x$. These are wavepackets obtained by appropriate superpositions of plane waves
\[
A_A = \epsilon_A(k,k_z) e^{i k \cdot x + i k_z \cdot z}
\]
So the meson sector of our field theory (at least, the part coming from the gauge field fluctuations) behaves more like a conformal field theory with a continuous spectrum than a massive field theory with particles.

It is interesting to note that the behavior we find depends crucially on the relative normalization of the Chern-Simons and Born-Infeld terms in the probe D4-brane action. If we change the relative coefficient even slightly, the function $R(y)$ changes its asymptotic behavior. In the effective quantum mechanics problem, the effective potential is still $V_A(z) = - \lambda$ but now $z_\infty$ is finite. Since the fluctuation must vanish at $z = \pm z_\infty$, the quantum mechanics problem now has a discrete spectrum (that of an infinite square well), and we would have a discrete spectrum of mesons in the dual field theory.

\subsection{Transverse scalar fluctuations}

For the transverse scalar fluctuations, the effective potential (\ref{veff}) has a lambda independent part and a term proportional to $\lambda$,  plotted in figure 7 and 8 respectively.\footnote{Note that we are plotting the potentials in this section as a function of $y$ rather than as a function of $z$, thus, the actual effective potential in the effective quantum mechanical problems will be related to the ones show by a reparametrization of the horizontal axis.}

\begin{figure}
\centering
\includegraphics[width=0.5\textwidth,angle=-90]{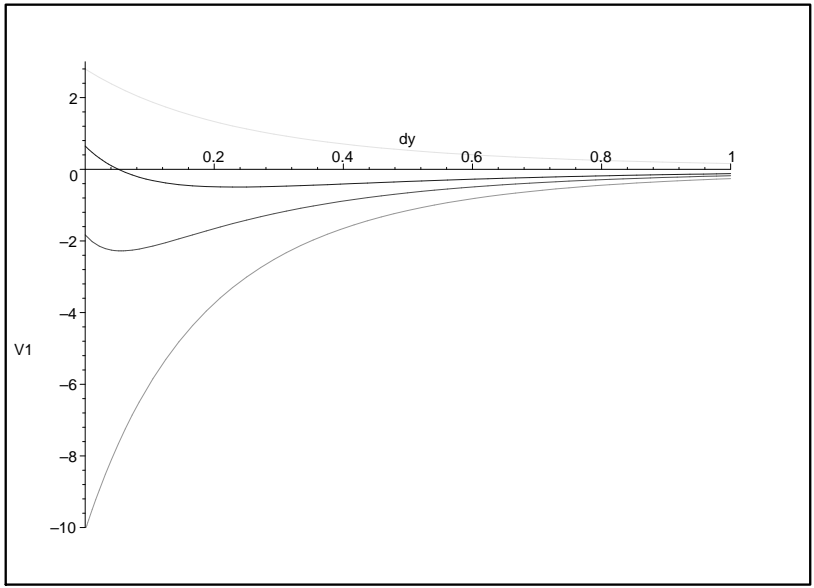}
\caption{Effective potential $V_1(y) = C(y)B(y)$ for $v_0=0$ and various values of $y_0$. Lower graphs have smaller $y_0$.}
\end{figure}

\begin{figure}
\centering
\includegraphics[width=0.5\textwidth,angle=-90]{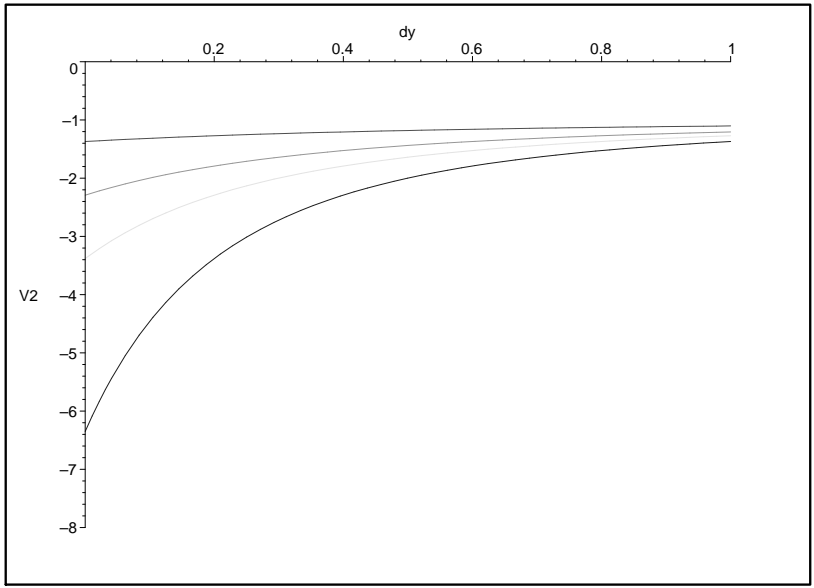}
\caption{Effective potential $V_2(y) = -A(y)B(y)$ for $v_0=0$ and various values of $y_0$. Lower graphs have smaller $y_0$.}
\end{figure}

For the $\lambda$-independent part of the potential $V_1 = B C$, we see that for large enough $y_0$ there will be no bound states, only a continuous spectrum with $E \ge 0$. For these values of $y_0$, there will be a zero energy (albeit non-normalizible) solution to the Schrodinger equation for all $\lambda \ge 0$ and no negative $\lambda$. Thus, the situation is similar to that for the gauge field modes.

For $y_0$ small enough, the potential $V_1$ will have one or more bound states with $E < 0$. For the full potential $V_1 + \lambda V_2$, these bound state energies increase as we decrease $\lambda$ below zero, and so we will have a bound state with zero energy for one or more negative values of $\lambda$. Thus, the field theory will be unstable for $y_0$ below some critical value $y_*(v_0)$ at which the potential $V_1$ develops a normalizible bound state. We anticipated this instability in the previous section as the tendency for certain brane configurations to ``slip'' over the sphere and lower their action. Based on that intuition, we expect that the critical value $y_*(v_0)$ will be the same value for which the asymptotic behavior of the brane configuration has $\theta_\infty = \pi/2$.

\subsection{$X_4$ and $\theta$ fluctuations.}

In general, the spectrum of fluctuations in the $X_4$ and $\theta$ directions is more complicated to obtain, since the equations are coupled, but we can analyze the fluctuations in the simple case where $v_0 = \pm \infty$. For $v_0  = -\infty$, the effective potentials for the $X_4$ and $\theta$ fluctuations are shown in figures 9 and 10.

\begin{figure}
\centering
\includegraphics[width=0.5\textwidth,angle=-90]{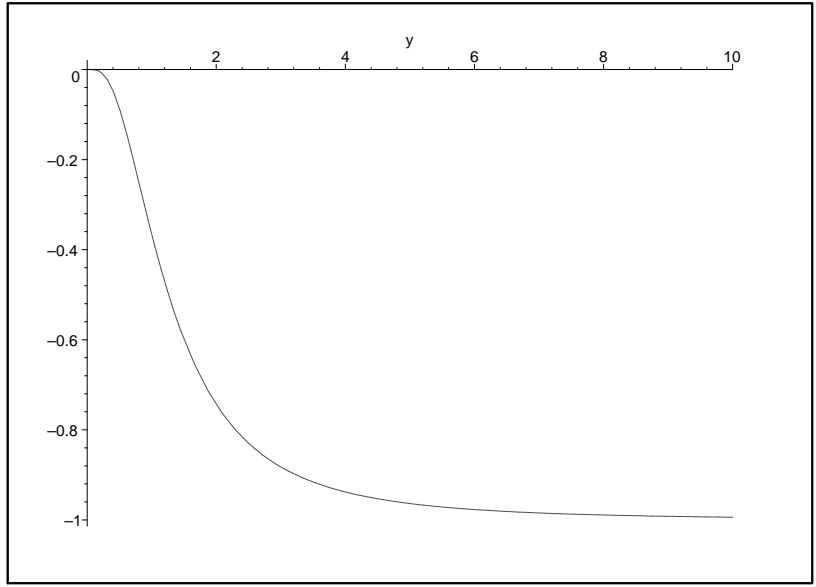}
\caption{Effective potential $V_4(y) = - A(y) B(y)$ for $v_0=-\infty$ and $y_0=1$. }
\end{figure}

\begin{figure}
\centering
\includegraphics[width=0.5\textwidth,angle=-90]{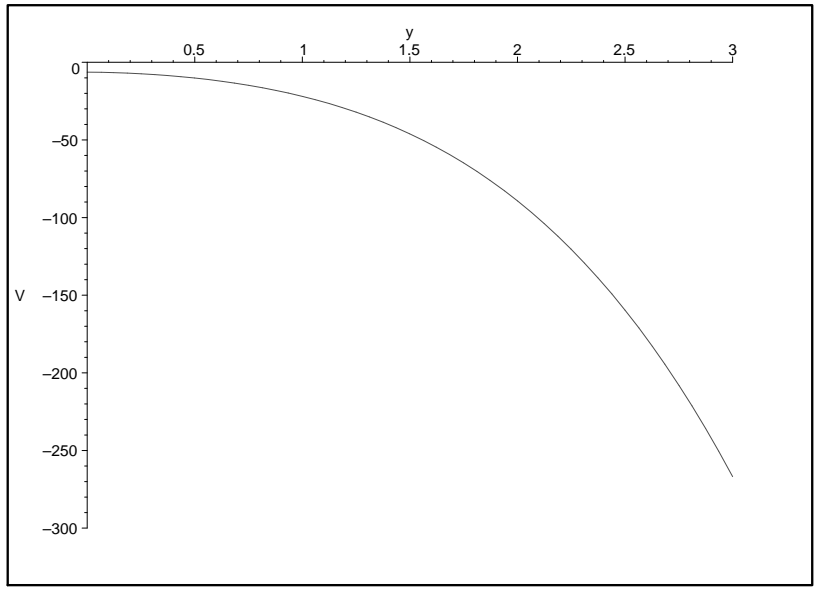}
\caption{Effective potential $V_\theta(y) = - A(y) B(y)$ for $v_0=-\infty$ and $y_0=1$. }
\end{figure}

Again, we multiply each of these by $\lambda$ and ask for which values of $\lambda$ the effective quantum mechanics problem has a zero energy eigenvalue. The result is again that any value $\lambda \ge 0$, but no negative values of $\lambda$ will work. Thus, at least for some values of parameters, we have shown that the model is tachyon-free. Since we do have modes for which the quadratic action is arbitrarily small, higher order terms (e.g. quartic in the fluctuations) may be important, but an analysis of these lies beyond the scope of the present work.

\section{Baryons}

By the gauge-theory / gravity dictionary, gauge fields in the bulk can be associated with conserved currents in the boundary theory. As in the Sakai-Sugimoto model, we have a gauge field living on the probe branes associated with our fundamental matter. This corresponds to the conserved baryon number (more precisely, quark number) current in the dual field theory. Specifically, the boundary value of the electrostatic potential $A_0$ (equivalently, the non-normalizible mode) corresponds to a chemical potential for baryon charge, while the electric flux at the boundary (equivalently, the normalizible mode of $A_0$) corresponds to the expectation value of baryon charge.

In order to have a state in the field theory with baryon charge, we need a source for electric flux on the probe brane. The simplest such source is a fundamental string endpoint (recall that the string action has a boundary term $\int A$ where $A$ is the gauge field on the brane). We can think of such an endpoint as corresponding to a single fundamental quark. If a string has both of its ends on the probe brane, we have two endpoints, but with opposite orientations, so this corresponds to a mesonic state with a quark and anti-quark. In order to have a baryon state, we must have $N$ strings with the same orientation ending on the probe brane. For a finite energy state, these strings must begin at some other source in the bulk. In our background, such a source is provided by D4-branes wrapped on $S^4$ \cite{wittenbaryon}. These necessarily have $N$ string endpoints, since the background D4-brane flux gives rise to $N$ units of charge on the spherical D4-branes, so we need $N$ units of the opposite charge (coming from the string endpoints) to satisfy the Gauss law constraint.

Thus, a finite energy configuration with a single unit of baryon charge is given by a D4-brane wrapped on $S^4$ together with $N$ fundamental strings stretched between this D4-brane and the probe D4-branes. A special feature of our model is that these wrapped D4-branes can smoothly combine with the probe D4-brane (after shrinking the strings to zero size) to give a configuration with lower energy. In the final configuration, there are no explicit fundamental strings; we simply have a configuration of the probe brane that now wraps the $S^4$. In the final configuration, the source for the electric field on the brane is the bulk flux of the Ramond-Ramond four-form, via the coupling $\int a \wedge F_4$.

Mathematically, the baryon charge in the field theory corresponds to an element of $\pi_4(S^4)  = \mathbb{Z}$ associated with the embedding. To see this, note that the probe brane embeddings correspond to mappings to the bulk space from $R^4$, topologically equivalent to a ball if we add the sphere at infinity. Given any probe brane embedding ${\cal E}$ with the same asymptotic behavior as the vacuum embedding ${\cal E}_0$, we can define a map from a topological $S^4$ to the bulk spacetime by splitting the $S^4$ into two balls along an $S^3$ and using the maps ${\cal E}$ and ${\cal E}_0$ to define the maps from the two balls. By considering only the sphere directions in the bulk, we can project this down to a mapping $S^4 \to S^4$, and such mappings may be associated with elements of the homotopy group $\pi_4(S^4) \sim \mathbb{Z}$. This integer gives the baryon number of the configuration in the field theory.

In order to find the actual bulk embedding corresponding to a single baryon, it is necessary to find the probe brane embedding with a single unit of winding on the $S^4$ (relative to the vacuum embedding) which has the minimum energy. Similarly, to find the bulk embedding corresponding to a nucleus with $n$ baryons, we want to find the minimal energy brane embedding with $n$ units of winding on the $S^4$. In general, a numerical analysis will be required, but it should be possible to obtain precise results for the masses of small nuclei.

\section*{Acknowledgements}

We would like to thank Andreas Karch, Moshe Rozali, Eric Zhitnitsky, and especially Ofer Aharony for helpful discussions and comments. This work has been supported in part by the Natural Sciences and Engineering Research Council of Canada, the Alfred P. Sloan Foundation, and the Canada Research Chairs programme.

\appendix

\section{Ramond-Ramond forms}

The results of our analysis depend crucially on the relative normalization of the Born-Infeld and Chern-Simons parts of the brane action. Since there is some variation in the literature for the normalization of the Ramond-Ramond fields in the Witten background, we include here a derivation of the correct result (consistent with a subset of previous papers). We use the fact that with the correct Ramond-Ramond flux, a D4-brane wrapped on $S^4$ should have induced $N$ units of charge, so that a configuration with $N$ string endpoints (of the correct orientation) on the wrapped D4-brane should have zero integrated charge and thus satisfy the Gauss Law constraint for the compact surface. The brane action is
\[
\mu_4 \int (2 \pi \alpha') A \wedge F_4
\]
while the action for each string endpoint is simply
\[
\int A_p
\]
where $A_p$ is the pullback of the gauge field to the worldline of the string endpoint. We know that the four form is constant on the sphere, i.e. we have $F_4 = C \epsilon_4$ where $\epsilon_4$ is the volume form on the sphere whose integral is $8 \pi^2 / 3$.

From this information, we find that the charge density on a wrapped D4-brane with string endpoints fixed at various locations $\Omega_i$ is
\[
\rho = \mu_4 (2 \pi \alpha') C \epsilon_4 - \sum_i \delta(\Omega - \Omega_i)
\]
where the delta functions are defined as four-forms localized at the indicated point that integrate to 1. Integrating the density over the sphere and setting the result to zero gives:
\[
\mu_4 (2 \pi \alpha') C \omega_4 = N
\]
Using the result that $\mu_4 = (2 \pi)^{-4} (\alpha')^{-5/2}$, we conclude that the Ramond-Ramond four-form is
\[
F_4 = 3 \pi N (\alpha')^{3 \over 2} \epsilon_4
\]
From this, we can calculate that the dual six-form in the background is given by (the sign here is related to a choice of convention for the direction of the flux)
\beas
(F_6)_{01234U} &=& - G_{00} G_{11} G_{22} G_{33} G_{44} G_{UU} {1 \over \sqrt{-G}} (F_4)_{\theta_1 \theta_2 \theta_3 \theta_4} \cr
&=& {U^2 \over F_4^6} 3 \pi N (\alpha')^{3 \over 2}
\eeas
and using $F_6 = dC_5$, we obtain the five-form
\[
C_5 = {U^3 - U_0^3 \over F_4^6}  \pi N (\alpha')^{3 \over 2} dt \wedge dx_1 \wedge dx_2 \wedge dx_3 \wedge dx_4 \; .
\]
where we have fixed the constant of integration so that the form is nonsingular at $U = U_0$.

\end{document}